\documentclass[12pt,letterpaper]{article}
\usepackage{epsfig,rotating,setspace,latexsym,amsmath,epsf,amssymb,bm}
\usepackage{cite,authblk}

\title{Secure Source Coding with a Helper\thanks{E-mail: tandonr@vt.edu, ulukus@umd.edu, kannanr@eecs.berkeley.edu.This work was supported by NSF Grants CCF $04$-$47613$, CCF $05$-$14846$, CNS $07$-$16311$ and CCF $07-29127$, and presented in part at the $47$th Annual Allerton Conference on Communications, Control and Computing, Monticello, IL, September $2009$.}}
\vspace{3cm}
\author[1]{Ravi Tandon}
\author[2]{Sennur Ulukus}
\author[3]{Kannan Ramchandran}
\affil[1]{\small Department of ECE, Virginia Tech, Blacksburg, VA, USA.}
\affil[2]{\small Department of ECE, University of Maryland, College Park, MD, USA.}
\affil[3]{\small Department of EECS, University of California, Berkeley, CA, USA.}

\newtheorem{Theo}{Theorem}

\newtheorem{Remark}{Remark}

\begin{document}
\maketitle
\thispagestyle{empty}

\begin{abstract}
We consider a secure lossless source coding problem with a  rate-limited helper. In particular,  Alice observes an independent and identically distributed (i.i.d.) source $X^{n}$ and wishes to transmit this source losslessly to Bob over a rate-limited link of capacity not exceeding $R_{x}$. A helper, say Helen, observes an i.i.d. correlated source $Y^{n}$ and can transmit information to Bob over another link of  capacity not exceeding $R_{y}$. A passive eavesdropper (say Eve) can observe the coded output of Alice, i.e., the link from Alice to Bob is public. The uncertainty about the source $X^{n}$ at Eve, (denoted by $\Delta$) is measured by the conditional entropy $\frac{H(X^{n}|J_{x})}{n}$, where $J_{x}$ is the coded output of Alice and $n$ is the block length. We completely characterize the rate-equivocation region for this secure source coding model, where we show that \textit{Slepian-Wolf binning} of $X$ with respect to the coded side information received at Bob is optimal. We next consider a modification of this model in which Alice also has access to the coded output of Helen. We call this model as the two-sided helper model. For the two-sided helper model, we characterize the rate-equivocation region. While the availability of side information at Alice does not reduce the rate of transmission from Alice, it significantly enhances the resulting equivocation at Eve. In particular, the resulting equivocation for the two-sided helper case is shown to be $\min(H(X),R_{y})$, i.e., one bit from the two-sided helper provides one bit of uncertainty at  Eve. From this result, we infer that Slepian-Wolf binning of $X$ is suboptimal and one can further decrease the information leakage to the eavesdropper by utilizing the side information at Alice. We finally generalize both of these results to the case in which there is additional uncoded side information $W^{n}$ available at Bob and characterize the rate-equivocation regions under the assumption that $Y^{n}\rightarrow X^{n} \rightarrow W^{n}$ forms a Markov chain.
\end{abstract}

\newpage
\section{Introduction}\label{intro}
The study of information theoretic secrecy was initiated by Shannon
in \cite{Shannon:1949}. Following Shannon's work, significant
contributions were made by Wyner \cite{WynerWiretap} who established
the rate-equivocation region of a degraded broadcast
channel. Wyner's result was generalized to the case of a general
broadcast channel by Csiszar and Korner \cite{CsiszarKorner}.
Recently, there has been a resurgence of activity in studying
multi-terminal and vector extensions of \cite{WynerWiretap},
\cite{CsiszarKorner}.

\begin{figure}[t]
\begin{minipage}[b]{0.5\linewidth}
\centering
\includegraphics[scale=0.63]{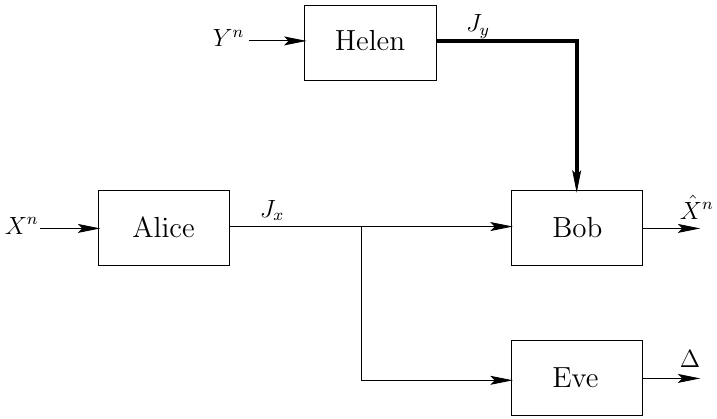}
\caption{One-sided helper.}
\label{fig:figure1}
\end{minipage}
\hspace{0.12cm}
\begin{minipage}[b]{0.5\linewidth}
\centering
\includegraphics[scale=0.63]{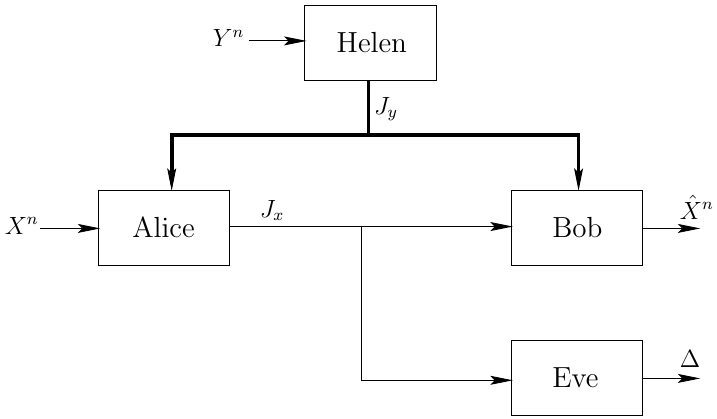}
\caption{Two-sided helper.}
\label{fig:figure2}
\end{minipage}
\end{figure}

In this paper, we investigate a secure transmission problem from a
source coding perspective. In particular, we first consider a
simple setup consisting of four terminals. Terminal $1$ (say
Alice) observes an i.i.d. source $X^{n}$ which it intends to
transmit losslessly to terminal $2$ (say Bob). A malicious but
passive user (say Eve) can observe the coded output of Alice.
In other words, the communication link between Alice and Bob is
public (or insecure). It is clear that since the malicious user
gets the same information as the legitimate user, there cannot be
any positive secret rate of transmission, i.e., some information about $X^{n}$ will be leaked to Eve. 
On the other hand, if there is a helper, say Helen,  who observes an i.i.d. source
$Y^{n}$ which is correlated with the source $X^{n}$ and transmits
information over a \textit{secure} rate-limited link to Bob, then one can
aim for creating uncertainty at the eavesdropper (see Figure
\ref{fig:figure1}\footnote{In Figures
  \ref{fig:figure1} and \ref{fig:figure2}, secure links are shown by {\bf{bold}} lines.}).
For the model shown in Figure \ref{fig:figure1}, we completely
characterize the rate-equivocation region. From our result, we
observe that the classical achievablity scheme of Ahlswede and
Korner \cite{AhlswedeKorner:1975} and Wyner \cite{Wyner:1975} for
source coding with rate-limited side information is robust in the
presence of a passive eavesdropper. By robust, we mean that in the presence of a passive adversary,
there is no need to change the original scheme as it achieves the maximum possible equivocation at Eve.

Next, we consider the model where Alice also has access to the
coded output of Helen and completely characterize the
rate-equivocation region. We will call this model the two-sided
helper model (see Figure \ref{fig:figure2}). From our result, we
observe that the availability of additional coded side information
at Alice allows her to increase uncertainty of the source at Eve
even though the rate needed by Alice to transmit the source
losslessly to Bob remains the same. This observation is in
contrast with the case of insecure source coding with side
information where providing coded side information to Alice is of
no value in terms of reducing Alice's transmission rate \cite{AhlswedeKorner:1975}.

We finally extend these results to the case in which there is additional \textit{uncoded} correlated side information  
$W^{n}$ available to Bob. We completely characterize the rate-equivocation region for this model when $Y^{n}\rightarrow X^{n}\rightarrow
W^{n}$ forms a Markov chain. We explicitly compute the rate-equivocation region for the cases of one-sided helper
and two-sided helper for a pair of binary symmetric sources. We show that having access to Helen's coded output at Alice yields a strictly larger
equivocation than the case of one-sided helper.
\newline\\
\textbf{Related Work}: The secure source coding setup shown in Figure $1$ was considered
in \cite{Deniz:ITW08} where it was also assumed that Eve has
access to additional correlated side information $Z^{n}$. Inner
and outer bounds for the rate-equivocation region were provided
for this setup, which do not match in general. The
rate-equivocation region was completely characterized in
\cite{Deniz:ITW08} for the case when Bob has complete uncoded side
information $Y^{n}$ and Eve has additional side information
$Z^{n}$. This result also follows from \cite{VinodKannan:ITW07}
where a similar three terminal setup was studied and the maximum
uncertainty at Eve was characterized under the assumption of no
rate constraint in the lossless transmission of the source to Bob.
A similar model was also studied in \cite{KundurG:2007} where Bob
intends to reconstruct both $X^{n}$ and $Y^{n}$ losslessly. It was
shown that Slepian-Wolf binning suffices for characterizing the
rate-equivocation region when the eavesdropper does not have
additional correlated side information. This setup was generalized
in \cite{Deniz:ISIT08} to the case when the eavesdropper has
additional side information $Z^{n}$, and inner and outer bounds
were provided, which do not match in general.

In \cite{Grokop:ISIT05}, a multi-receiver secure broadcasting
problem was studied, where Alice intends to transmit a source
$X^{n}$ to $K$ legitimate users. The $k$th user has access to a
correlated source $Y_{k}^{n}$, where $Y_{k}^{n}=X^{n}\oplus
B_{k}^{n}$, for $k=1,\ldots K$, and the eavesdropper has access to
$Z^{n}$, where $Z^{n}=X^{n}\oplus E^{n}$, and the noise sequences
$(B_{1}^{n},\ldots, B_{K}^{n},E^{n})$ are mutually independent and
also independent of the source $X^{n}$. Furthermore, it was
assumed that Alice also has access to $(Y_{1}^{n},\ldots
Y_{K}^{n})$. For sources with such modulo-additive structure, it
was shown that to maximize the uncertainty at the eavesdropper,
Alice cannot do any better than describing the error sequences
$(B_{1}^{n},\ldots,B_{K}^{n})$ to the legitimate users. This model
is related to the two-sided helper model shown in Figure
\ref{fig:figure2}; see Section \ref{twosidedH} for details.
%
\newline\\
\textbf{Summary of Main Results}: In Section \ref{onesidedH}, we present the rate-equivocation
region for the case of one-sided helper. We show that Slepian-Wolf
binning alone at Alice is optimal for this case. We present the
rate-equivocation region for the case of two-sided helper in
Section \ref{twosidedH}.  For the case of two-sided helper, Alice
utilizes the coded-side information received from Helen as follows: it can 
narrow down the set of uncertainty about $X$-sequences at Bob given the output received
from Helen. It only sends the residual information necessary to decode $X^{n}$ at Bob. 
We show that the resulting equivocation of this scheme is $\min(H(X),R_{y})$, i.e., one secure 
(two-sided) bit from Helen results in one bit of equivocation at Eve. 
From this result, we demonstrate the insufficiency of Slepian-Wolf binning at Alice by explicitly utilizing the side
information at Alice. This observation is further highlighted in
Section \ref{Sec:Comparison} where we compare the rate-equivocation
regions of two-sided helper and one-sided helper cases for a pair
of binary symmetric sources. For this example, we show that for
all $R_{y}>0$, the information leakage to the eavesdropper for the
two-sided helper is strictly less than the case of one-sided
helper. We finally generalize these results to the case when there is additional side informations
$W$ at Bob. For the case in which $Y\rightarrow X \rightarrow W$, we characterize the tradeoff of rates and equivocation.
For the case of two-sided helper, the optimal resulting equivocation at Eve is $\min(H(X), R_{y}+I(X;W))$, i.e., 
the net equivocation resulting from coded and uncoded side information is \textit{additive} in nature. By additive we mean the following:
suppose that $W$ was not present, then the equivocation would be $\min(H(X),R_{y})$ from our result of two-sided helper. On the other hand, if $R_{y}=0$, then we know from \cite{VinodKannan:ITW07}, that the optimal equivocation is given by $I(X;W)$. Thus, in the presence of both uncoded and coded side-information, the net equivocation is $R_{y}+I(X;W)$ till it saturates to $H(X)$. Parts of this paper have been presented in \cite{TandonAllerton2009}.

\section{Main Results}
\subsection{One-Sided Helper}\label{onesidedH}
We consider the following source coding problem. Alice observes an
$n$-length source sequence $X^{n}$, which is intended to be
transmitted losslessly to Bob. The coded output of Alice can be
observed by the malicious user Eve. Moreover, Helen observes a
correlated source $Y^{n}$ and there exists a noiseless
rate-limited channel from Helen to Bob. We assume that the link
from Helen to Bob is a secure link and the coded output of Helen
is not observed by Eve (see Figure $1$). The sources
$(X^{n},Y^{n})$ are generated i.i.d. according to $p(x,y)$ where
$p(x,y)$ is defined over the finite product alphabet
$\mathcal{X}\times \mathcal{Y}$. The aim of Alice is to create
maximum uncertainty at Eve regarding the source $X^{n}$ while
losslessly transmitting the source to Bob.


An $(n, 2^{nR_{x}},2^{nR_{y}})$ code for this model consists of an
encoding function at Alice, $f_{x}:X^{n}\rightarrow \{1,\ldots,2^{nR_{x}}\}$,
an encoding function at Helen, $f_{y}:Y^{n}\rightarrow \{1,\ldots,2^{nR_{y}}\}$, and a decoding function at Bob, $g: \{1,\ldots,2^{nR_{x}}\} \times \{1,\ldots,2^{nR_{y}}\} \rightarrow X^{n}$.
The uncertainty about the source $X^{n}$ at Eve is measured by
$H(X^{n}|f_{x}(X^{n}))/n$. The probability of error in the reconstruction of
$X^{n}$ at Bob is defined as
$P_{e}^{n}=\mbox{Pr}(g(f_{x}(X^{n}),f_{y}(Y^{n}))\neq X^{n})$.
A triple $(R_{x},R_{y},\Delta)$ is achievable if for any $\epsilon>0$,
there exists a $(n, 2^{nR_{x}}, 2^{nR_{y}})$ code such that $P_{e}^{n}\leq
\epsilon$ and \break $H(X^{n}|f_{x}(X^{n}))/n \geq \Delta$. We denote
the set of all achievable $(R_{x},R_{y},\Delta)$ rate triples as $\mathcal{R}_{1-sided}$.

The main result is given in the following theorem.
\begin{Theo}\label{Theorem1} The set of achievable rate triples $\mathcal{R}_{1-sided}$ for
  secure source coding with one-sided helper is given as
\begin{align}
\mathcal{R}_{1-sided}=\Big\{(R_{x},R_{y},\Delta): R_{x}&\geq H(X|V)\\
R_{y}&\geq I(Y;V) \\
\Delta&\leq I(X;V)\Big\}
\end{align}
where the joint distribution of the involved random variables is as
follows,
\begin{align}
p(x,y,v)&=p(x,y)p(v|y)
\end{align}
and it suffices to consider such distributions for which
$|\mathcal{V}|\leq |\mathcal{Y}|+2$.
\end{Theo}
The proof of Theorem \ref{Theorem1} is given in the Appendix.

We note that inner and outer bounds for source coding model considered 
in this section were presented in \cite[Theorem $3.1$]{Deniz:ITW08} although these
bounds do not match in general. These bounds match when Bob has complete 
uncoded side information $Y^{n}$, i.e., when $R_{y}\geq H(Y)$.

The achievability scheme which yields the rate region described in Theorem \ref{Theorem1} is summarized as follows:
\begin{enumerate}
\item Helen describes the source $Y^{n}$ to Bob through a coded
  output $V^{n}$.
\item Alice performs
Slepian-Wolf binning of the source $X^{n}$ with respect to the coded
side information, $V^{n}$, available at Bob.
\end{enumerate}
Therefore, this result shows that the achievable scheme of
Ahlswede, Korner \cite{AhlswedeKorner:1975} and Wyner
\cite{Wyner:1975} is optimal in the presence of an eavesdropper.
Moreover, upon dropping the security constraint, Theorem
\ref{Theorem1} yields the result of
\cite{AhlswedeKorner:1975},\cite{Wyner:1975}.

\subsection{Two-Sided Helper}\label{twosidedH}

We next consider the following modification of the model
considered in Section \ref{onesidedH}. In this model, Alice also
has access to the coded output of Helen besides the source
sequence $X^{n}$ (see Figure $2$). An $(n, 2^{nR_{x}},2^{nR_{y}})$
code for this model consists of an encoding function at Alice,
$f_{x}:X^{n} \times \{1,\ldots,2^{nR_{y}}\} \rightarrow
\{1,\ldots,2^{nR_{x}}\}$, an encoding function at Helen,
$f_{y}:Y^{n}\rightarrow \{1,\ldots,2^{nR_{y}}\}$, and a decoding
function at Bob, $g: \{1,\ldots,2^{nR_{x}}\} \times
\{1,\ldots,2^{nR_{y}}\} \rightarrow X^{n}$. The uncertainty about
the source $X^{n}$ at Eve is measured by
$H(X^{n}|f_{x}(X^{n}))/n$. The probability of error in the
reconstruction of $X^{n}$ at Bob is defined as
$P_{e}^{n}=\mbox{Pr}(g(f_{x}(X^{n},f_{y}(Y^{n})),f_{y}(Y^{n}))\neq
X^{n})$. A triple $(R_{x},R_{y},\Delta)$ is achievable if for any
$\epsilon>0$, there exists a $(n, 2^{nR_{x}}, 2^{nR_{y}})$ code
such that $P_{e}^{n}\leq \epsilon$ and $H(X^{n}|f_{x}(X^{n}))/n
\geq \Delta$. We denote the set of all achievable
$(R_{x},R_{y},\Delta)$ rate triples as $\mathcal{R}_{2-sided}$.

The main result is given in the following theorem.

\begin{Theo}\label{Theorem2} The set of achievable rate triples $\mathcal{R}_{2-sided}$ for
secure source coding with two-sided helper is given as
\begin{align}
\mathcal{R}_{2-sided}=\Big\{(R_{x},R_{y},\Delta): R_{x}&\geq H(X|V)\\
\vspace{-0.1in}R_{y}&\geq I(Y;V)\\
\Delta&\leq \min(H(X),R_{y})\Big\}\label{Theo2}
\end{align}
where the joint distribution of the involved random variables is as
follows,
\begin{align}
p(x,y,v)&=p(x,y)p(v|y)
\end{align}
and it suffices to consider such distributions for which
$|\mathcal{V}|\leq |\mathcal{Y}|+2$.
\end{Theo}
The proof of Theorem \ref{Theorem2} is given in the Appendix. 

The achievability scheme which yields the rate region described in
Theorem \ref{Theorem2} is summarized as follows:
\begin{enumerate}
\item Helen describes the source $Y^{n}$ to both Bob and
Alice through a coded output $V^{n}$.
\item Given the coded output $V^{n}$, Alice can narrow down the set of conditionally 
typical $X^{n}$-sequences, which are approximately $2^{nH(X|V)}$. Furthermore, 
for $n$ sufficiently large,  the observed $x^{n}$-sequence would belong to this set with high probability.
Alice sends the index of the observed sequence corresponding to the conditionally typical set for the received coded output. 
\end{enumerate}
Therefore, the main difference between the achievability schemes
for Theorems \ref{Theorem1} and \ref{Theorem2} is at the encoding
at Alice. Our encoding scheme at Alice for the case of two-sided helper comprises of the following key step: using the
coded side information and the source sequence, Alice narrows down the uncertainty at Bob by considering the set of typical $X$-sequences given the coded output from Helen. It then transmits the index to which the observed $X^{n}$-sequence falls in this set. The key observation is that the helper's output is two-sided and \textit{secure} (i.e., only available at Alice and Bob), and Eve only gets to observe the index of the $X$ sequence sent by Alice. Without any knowledge of the $V^{n}$-sequence, from Eve's point of view, the correct $X^{n}$-sequence could have resulted from any of the $2^{nR_{y}}$ conditionally typical sets, each corresponding to the total number of $V^{n}$-sequences; and thus the resulting equivocation at Eve is $\min(H(X),R_{y})$.

\begin{Remark}
Besides reflecting the fact that the uncertainty at Eve can be strictly larger than the case of a one-sided helper, Theorem
\ref{Theorem2} has another interesting interpretation. If Alice and Helen can use sufficiently large rates to securely transmit
the source $X^{n}$ to Bob, then the helper can simply transmit a secret key of entropy $H(X)$ to both Alice and Bob. Alice can then
use this secret key to losslessly transmit the source to Bob in perfect secrecy by using a one-time pad \cite{Shannon:1949}. In
other words, when $R_{x}$ and $R_{y}$ are larger than $H(X)$, one can immediately obtain this result from Theorem \ref{Theorem2} by
selecting $V$ to be independent of $(X,Y)$ and uniformly distributed on $\{1,\ldots,|\mathcal{X}|\}$. Perhaps the most interesting aspect of the result in Theorem \ref{Theorem2} is that for an arbitrary $R_{y}$, the two-sided coded output $V$ plays the dual role of providing security and reducing rate of transmission from Alice.
\end{Remark}

\begin{Remark}
Now consider the model where the side information $Y^{n}$ is
of the form $Y^{n}=X^{n}\oplus B^{n}$, where
$|\mathcal{B}|=|\mathcal{X}|$, and $B^{n}$ is independent of $X^{n}$.
Moreover, assume that the side information $Y^{n}$ is available to both Alice and
Bob in an \textit{uncoded} manner. For this model, it follows from \cite{Grokop:ISIT05}
that, to maximize the uncertainty at the eavesdropper, Alice cannot
do any better than describing the error sequence $B^{n}$ to Bob.
Note that our two-sided helper model differs from this model in two
aspects: first, in our case, the common side information available to
Alice and Bob is \textit{coded} and rate-limited, secondly, the sources in our
model do not have to be in modulo-additive form.
\end{Remark}

\subsection{Additional Uncoded Side Information at Bob}
We next present extensions of Theorems \ref{Theorem1} and \ref{Theorem2} to the case in which Bob has additional 
correlated side information $W^{n}$, and  we assume that $Y\rightarrow X\rightarrow W$ forms a Markov chain.

\begin{Theo}\label{Theorem1B} The set of achievable rate triples $\mathcal{R}^{W}_{1-sided}$ for
secure source coding with one-sided helper and side information $W$ at Bob is given as
\begin{align}
\mathcal{R}^{W}_{1-sided}=\Big\{(R_{x},R_{y},\Delta): R_{x}&\geq H(X|W,V)\\
\vspace{-0.1in}R_{y}&\geq I(Y;V|W)\\
\Delta&\leq I(X;V,W)\Big\}\label{Theo1B}
\end{align}
where the joint distribution of the involved random variables is as
follows,
\begin{align}
p(x,w,y,v)&=p(x,w)p(y|x)p(v|y)
\end{align}
and it suffices to consider such distributions for which
$|\mathcal{V}|\leq |\mathcal{Y}|+3$.
\end{Theo}

\begin{Theo}\label{Theorem3} The set of achievable rate triples $\mathcal{R}^{W}_{2-sided}$ for
secure source coding with two-sided helper and side information $W$ at Bob is given as
\begin{align}
\mathcal{R}^{W}_{2-sided}=\Big\{(R_{x},R_{y},\Delta): R_{x}&\geq H(X|W,V)\\
\vspace{-0.1in}R_{y}&\geq I(Y;V|W)\\
\Delta&\leq \min(H(X),R_{y}+I(X;W))\Big\}\label{Theo3}
\end{align}
where the joint distribution of the involved random variables is as
follows,
\begin{align}
p(x,w,y,v)&=p(x,w)p(y|x)p(v|y)
\end{align}
and it suffices to consider such distributions for which
$|\mathcal{V}|\leq |\mathcal{Y}|+3$.
\end{Theo}
The proofs of Theorems \ref{Theorem1B} and \ref{Theorem3} are given in the Appendix. 

\section{Example: Binary Symmetric Sources}\label{Sec:Comparison}
In this section, we compare the rate-equivocation tradeoffs presented in Theorems \ref{Theorem1} and \ref{Theorem2} for a pair of binary sources.

Let $X$ and $Y$ be binary sources with $X\sim \mbox{Ber}(1/2),
Y\sim \mbox{Ber}(1/2)$ and $X=Y\oplus E$, where $E\sim
\mbox{Ber}(\delta)$. For this pair of sources, the region
described in Theorem \ref{Theorem1} can be completely
characterized as,
\begin{align}
\mathcal{R}_{1-sided}(R_{y})=\big\{(R_{x},\Delta):\hspace{0.05in}R_{x}&\geq h(\delta*h^{-1}(1-R_{y}))\nonumber\\
\Delta&\leq 1-h(\delta*h^{-1}(1-R_{y}))\big\}\label{regn1}
\end{align}
and the region in Theorem \ref{Theorem2} can be completely
characterized as,
\begin{align}
\mathcal{R}_{2-sided}(R_{y})=\big\{(R_{x},\Delta):\hspace{0.05in}R_{x}&\geq h(\delta*h^{-1}(1-R_{y}))\nonumber\\
\Delta&\leq \min(R_{y},1)\big\}
\end{align}
where $h(.)$ is the binary entropy function, and $a*b=a(1-b)+b(1-a)$.

We start with the derivation of (\ref{regn1}). Without loss of
generality, we assume that $R_{y}\leq H(Y)$. Achievability follows by
selecting $V=Y\oplus N$, where $N\sim \mbox{Ber}(\alpha)$, where
\begin{align}
\alpha&=h^{-1}(1-R_{y})
\end{align}
Substituting, we obtain
\begin{align}
H(X|V)&= h(\delta*h^{-1}(1-R_{y}))\\
I(X;V)&= 1-h(\delta*h^{-1}(1-R_{y}))
\end{align}
which completes the achievability. Note that $Y$ is independent of
$E$, and the random variables $X$, $Y$, and $V$ form a Markov
chain, i.e., $X\rightarrow Y\rightarrow V$. Using this Markov
chain, the converse follows by simple application of Mrs. Gerber's
lemma \cite{Gerber:1973} as follows. Let us be given $R_{y}\in
(0,1)$. We have
\begin{align}
R_{y}&\geq I(Y;V)\\
&= H(Y)-H(Y|V)\\
&= 1 - H(Y|V)
\end{align}
which implies $H(Y|V)\geq 1-R_{y}$. Mrs. Gerber's lemma states that for
$X=Y\oplus E$, with $E\sim \mbox{Ber}(\delta)$, if $H(Y|V)\geq \beta$, then $H(X|V)\geq
h(\delta*h^{-1}(\beta))$. We therefore have,
\begin{align}
R_{x}&\geq H(X|V)\\
&\geq h(\delta*h^{-1}(1-R_{y}))
\end{align}
and
\begin{align}
\Delta&\leq I(X;V)\\
&= H(X)-H(X|V)\\
&= 1- H(X|V)\\
&\leq 1- h(\delta*h^{-1}(1-R_{y}))
\end{align}
This completes the converse.

\begin{figure}[t]
\centering
\includegraphics[scale=0.53]{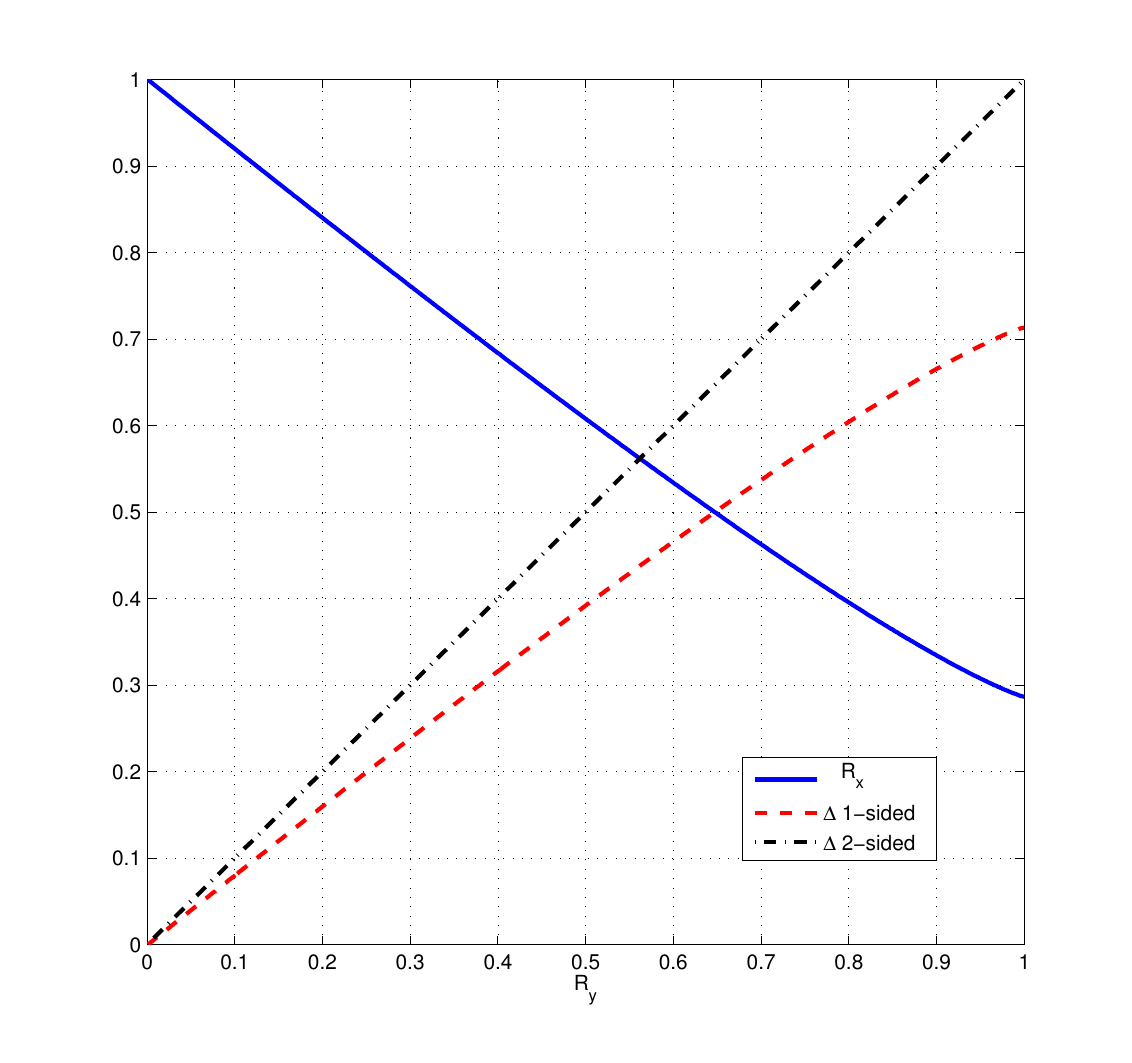}
   \caption{The rate-equivocation region for a pair of binary symmetric sources.}\label{fig4}
\end{figure}

The rate from Alice, $R_{x}$ and the equivocation $\Delta$ for the
cases of one-sided and two-sided helper are shown in Figure
\ref{fig4} for the case when $\delta=0.05$. For the one-sided
helper, we can observe a trade-off in the amount of information
Alice needs to send versus the uncertainty at Eve. For small
values of $R_{y}$, Alice needs to send more information thereby
leaking out more information to Eve. The amount of information
leaked (i.e, $I(X;V)= H(X)-\Delta$) has a one to one relationship to the information sent by Alice. On the other
hand, for the case of two-sided helper, the uncertainty at the
eavesdropper is always strictly larger than the uncertainty in the
one-sided case. Also note that for this pair of sources, perfect
secrecy is possible for the case of two-sided helper when
$R_{y}\geq H(Y)$ which is not possible for the case of one-sided
helper.

\section{Conclusions}
In this paper, we considered several secure source coding problems. We
first provided the characterization of the rate-equivocation region
for a secure source coding problem with coded side information at
the legitimate user. We next extended this result to the case in which the helper is two-sided, i.e., 
its output is available at both Alice and Bob. We characterized the
 rate-equivocation region for the case of two-sided helper.
The value of two-sided coded side information is emphasized by
comparing the respective equivocations for a pair of binary
sources. It is shown that Slepian-Wolf binning
alone is insufficient and using our achievable scheme, one attains
strictly larger uncertainty at the eavesdropper than the case of
one-sided helper. Finally, these results are extended to the case in which Bob has access to additional uncoded side information $W$.
Under the assumption that $Y\rightarrow X\rightarrow W$ forms a Markov chain, the rate-equivocation tradeoffs have been characterized
for both one-sided and two-sided scenarios.

\section{Appendix}
\subsection{Proof of Theorem \ref{Theorem1}}

\subsubsection{Achievability}
Fix the distribution $p(x,y,v)=p(x,y)p(v|y)$.
\begin{enumerate}
\item Codebook generation at Helen: From the conditional
probability distribution $p(v|y)$ compute
$p(v)=\sum_{y}p(y)p(v|y)$. Generate $2^{nR_{y}}$ codewords $v(l)$
independently according to $\prod_{i=1}^{n}p(v_{i})$, where
$l=1,\ldots,2^{nR_{y}}$.

\item Codebook generation at Alice: Randomly bin the $x^{n}$
sequences into $2^{nH(X|V)}$ bins and index these bins as
$m=1,\ldots,M$, where $M=2^{nH(X|V)}$.

\item Encoding at Helen: On observing the sequence $y^{n}$, Helen
tries to find a sequence $v(l)$ such that $(v(l),y^{n})$ are
jointly typical. From rate-distortion theory, we know that there
exists one such sequence as long as $R_{y}\geq I(V;Y)$. Helen
sends the index $l$ of the sequence $v(l)$.

\item Encoding at Alice: On observing the sequence $x^{n}$, Alice
finds the bin index $m_{X}$ in which the sequence $x^{n}$ falls
and transmits the bin index $m_{X}$.

\item Decoding at Bob: On receiving $l$ and the bin index $m_{X}$,
Bob tries to find a unique $x^{n}$ sequence in bin $m_{X}$ such
that $(v(l),x^{n})$ are jointly typical. This is possible since
the number of $x^{n}$ sequences in each bin is roughly
$2^{nH(X)}/2^{nH(X|V)}$ which is $2^{nI(X;V)}$. The existence
of an $x^{n}$ such that $(v(l),x^{n})$ are jointly typical
is guaranteed by the Markov lemma \cite{Cover:book} and the uniqueness
is guaranteed by the properties of jointly typical sequences \cite{Cover:book}. \item
Equivocation:
\begin{align}
H(X^{n}|m_{X})&=H(X^{n},m_{X})-H(m_{X})\\
&=H(X^{n})+H(m_{X}|X^{n})-H(m_{X})\\
&= H(X^{n})-H(m_{X})\\
&\geq H(X^{n})-\mbox{log}(M)\\
&=H(X^{n})-nH(X|V)\\
&= n I(X;V)
\end{align}
Therefore,
\begin{align}
\Delta&\leq I(X;V)
\end{align}
is achievable. This completes the achievability part.
\end{enumerate}

\subsubsection{Converse}\label{subsubcite}
Let the output of the helper be $J_{y}$, and the
output of Alice be $J_{x}$, i.e.,
\begin{align}
J_{y}&=f_{y}(Y^{n})\\
J_{x}&=f_{x}(X^{n})
\end{align}
First note that, for noiseless reconstruction of the sequence $X^{n}$ at the legitimate decoder, we have by Fano's inequality
\begin{align}
H(X^{n}|J_{x},J_{y})&\leq n\epsilon_{n}\label{fano}
\end{align}

We start by obtaining a lower bound on $R_{x}$, the rate of Alice, as follows
\begin{align}
nR_{x}&\geq H(J_{x})\\
&\geq H(J_{x}|J_{y})\\
&= H(X^{n},J_{x}|J_{y})-H(X^{n}|J_{x},J_{y})\\
&\geq H(X^{n},J_{x}|J_{y})-n\epsilon_{n}\label{e1}\\
&\geq H(X^{n}|J_{y})-n\epsilon_{n}\\
&=\sum_{i=1}^{n}H(X_{i}|X^{i-1},J_{y})-n\epsilon_{n}\\
&=\sum_{i=1}^{n}H(X_{i}|V_{i})-n\epsilon_{n}\label{e2}\\
&= nH(X_{Q}|V_{Q},Q)-n\epsilon_{n}\\
&=nH(X|V)-n\epsilon_{n}\label{etemp2}
\end{align}
where (\ref{e1}) follows by (\ref{fano}). In (\ref{e2}), we have defined
\begin{align}
V_{i}&=(J_{y},X^{i-1})\label{defnV}
\end{align}
In (\ref{etemp2}), we have defined,
\begin{align}
X=X_{Q}, \quad V=(Q,V_{Q})
\end{align}
where $Q$ is uniformly distributed on $\{1,\ldots,n\}$ and is independent of all
other random variables.

Next, we obtain a lower bound on $R_{y}$, the rate of the helper,
\begin{align}
nR_{y}&\geq H(J_{y})\\
&\geq I(J_{y};Y^{n})\\
&= \sum_{i=1}^{n}I(J_{y},Y^{i-1};Y_{i})\\
&= \sum_{i=1}^{n}I(J_{y},Y^{i-1},X^{i-1};Y_{i})\label{e3}\\
&\geq \sum_{i=1}^{n}I(J_{y},X^{i-1};Y_{i})\\
&= \sum_{i=1}^{n}I(V_{i};Y_{i})
\end{align}
\begin{align}
&= nI(V_{Q};Y_{Q}|Q)\\
&=nI(V;Y)\label{etemp3}
\end{align}
where (\ref{e3}) follows from the Markov chain
\begin{align}
X^{i-1}\rightarrow (J_{y},Y^{i-1}) \rightarrow Y_{i}
\end{align}
and in (\ref{etemp3}), we have defined $Y=Y_{Q}$.

We now have the main step, i.e., an upper bound on the equivocation
rate of the eavesdropper,
\begin{align}
H(X^{n}|J_{x})&=
H(X^{n},J_{y}|J_{x})-H(J_{y}|X^{n},J_{x})\\
&=
H(J_{y}|J_{x})-H(J_{y}|X^{n},J_{x})+H(X^{n}|J_{x},J_{y})\\
&=
H(J_{y}|J_{x})-H(J_{y}|X^{n})+H(X^{n}|J_{x},J_{y})\label{e4}\\
&\leq
H(J_{y})-H(J_{y}|X^{n})+H(X^{n}|J_{x},J_{y})\\
&\leq I(J_{y};X^{n}) + n\epsilon_{n}\label{e5}\\
&= \sum_{i=1}^{n}I(J_{y};X_{i}|X^{i-1}) + n\epsilon_{n}\\
&= \sum_{i=1}^{n}I(J_{y},X^{i-1};X_{i}) +
n\epsilon_{n}\\
&=\sum_{i=1}^{n}I(X_{i};V_{i})+
n\epsilon_{n}\\
&= nI(X_{Q};V_{Q}|Q)+n\epsilon_{n}\\
&= nI(X;V)+n\epsilon_{n}
\end{align}
where (\ref{e4}) follows from the Markov chain
\begin{align}
J_{x}\rightarrow X^{n} \rightarrow J_{y}
\end{align}
and (\ref{e5}) follows from (\ref{fano}). This implies
\begin{align}
\Delta&\leq I(X;V)
\end{align}

Also note that the following is a Markov chain,
\begin{align}
V\rightarrow Y \rightarrow X
\end{align}
Therefore, the joint distribution of the involved random variables is
\begin{align}
p(x,y,v)&=p(x,y)p(v|y)
\end{align}
From support lemma \cite{Csiszar:book}, it can be shown that it
suffices to consider such joint distributions for which
$|\mathcal{V}|\leq |\mathcal{Y}|+2$.

In (\ref{defnV}), we have defined the auxiliary random variable as
$V_{i}=(J_{y},X^{i-1})$. We remark here that the converse for
Theorem \ref{Theorem1} can also be proved by defining,
$V_{i}=(J_{y},Y^{i-1})$ as in \cite[Section $14.8$]{Cover:book}.
Note that due to the fact that the sources $(X^{n},Y^{n})$ are
generated in an i.i.d. manner, the following is a Markov chain,
\begin{align}
(J_{y},Y^{i-1},X^{i-1})\rightarrow Y_{i}\rightarrow X_{i}\label{longMC}
\end{align}
This is due to the fact that $X_{i}$ does not carry any extra
information about \break $(J_{y}=f_{y}(Y^{n}),Y^{i-1},X^{i-1})$
that is not there in $Y_{i}$. Therefore, (\ref{longMC}) implies
that the following are also valid Markov chains,
\begin{align}
&(J_{y},X^{i-1})\rightarrow Y_{i}\rightarrow X_{i}\\
&(J_{y},Y^{i-1})\rightarrow Y_{i}\rightarrow X_{i}
\end{align}
and the converse for Theorem \ref{Theorem1} can be proved by
defining $V_{i}=(J_{y},X^{i-1})$ or $V_{i}=(J_{y},Y^{i-1})$.

\subsection{Proof of Theorem \ref{Theorem2}}
\subsubsection{Achievability}
Fix the distribution $p(x,y,v)=p(x,y)p(v|y)$.
\begin{enumerate}
\item Codebook generation at Helen: From the conditional
probability distribution $p(v|y)$ compute
$p(v)=\sum_{y}p(y)p(v|y)$. Generate $2^{nR_{y}}$ codewords $v(l)$
independently according to $\prod_{i=1}^{n}p(v_{i})$, where
$l=1,\ldots,2^{nR_{y}}$.

\item Encoding at Helen: On observing the sequence $y^{n}$, Helen
tries to find a sequence $v(l)$ such that $(v(l),y^{n})$ are
jointly typical. If there exists such a sequence $v(l)$, it sends the index $l$ to Alice and Bob, otherwise it sends a fixed index $l=0$. 

\item Encoding at Alice: The key difference from the one-sided
helper case is in the encoding at Alice.  Let $\mathcal{E}_{H}=1$ denote the event that the encoding at Helen succeeds, i.e., there exists at least one $l$
such that $(v(l),y)\in T^{n}_{YV}$. The probability of this event can be made arbitrarily close to $1$, for $n$ sufficiently large as long as $R_{y}\geq I(Y;V)$. If $\mathcal{E}_{H}=1$, Alice receives the index $l$ of the sequence $v(l)$, otherwise it receives the fixed index $l=0$. 

Conditioned on the event $\mathcal{E}_{H}=1$, we note the following:

\begin{itemize}
\item $P(L=l|\mathcal{E}_{H}=1)\approx 2^{-nR_{y}}$, for $l=1,\ldots, 2^{nR_{y}}$, i.e., any of the $L$ indices are approximately equally likely\footnote{Formally, by the notation $P(A=a)\approx 2^{-nR}$, we refer to the following: \break $P(A=a)\in [2^{-n(R+\delta_{n})},2^{-n(R-\delta_{n})}]$, for some sequence $\delta_{n}$ such that $\delta_{n}\rightarrow 0$ as $n\rightarrow \infty$.}  to be sent given $\mathcal{E}_{H}=1$ for $n$ sufficiently large.
\item For each possible sequence $v(l)$ received from Helen, and given that $(v(l),y^{n})\in T^{n}_{YV}$, we denote the set of
of conditional typical $X$-sequences given $v(l)$ as $T^{n}_{X|v(l)}$, for $l=1,\ldots,2^{nR_{y}}$. 
\item From Markov lemma, we have that $P((X^{n},v(l))\in T_{X|v(l)}|\mathcal{E}_{H}=1, L=l)\geq 1-\epsilon_{n}$, where $\epsilon_{n}\rightarrow 0$ as $n\rightarrow \infty$, i.e., the observed $x^{n}$ sequence at Alice will belong to the conditional typical set $T^{n}_{X|v(l)}$ with high probability. 
\item For $n$ sufficiently large, we have $|T^{n}_{X|v(l)}|\approx 2^{nH(X|V)}$. Enumerate the sequences as $j=1,\ldots, 2^{nH(X|V)}$. 
\item The set of $x$-sequences belonging to $T^{n}_{X|v(l)}$ are approximately uniformly distributed, i.e., $P(X^{n}=x^{n}|X^{n}\in T^{n}_{X|v(l)})\approx 2^{-nH(X|V)}$.
\item For any $l\neq l^{'}$, the sets $T^{n}_{X|v(l)}$ and $T^{n}_{X|v(l^{'})}$ are disjoint, i.e., $|T^{n}_{X|v(l)} \cap T^{n}_{X|v(l^{'})}| \leq \epsilon_{n}$, where $\epsilon_{n}\rightarrow 0$ as $n\rightarrow \infty$. 
\end{itemize}
On observing the sequence $x^{n}$ and obtaining $v(l)$ from Helen, 
Alice sends the index $j$ corresponding to the conditionally typical set $T^{n}_{X|v(l)}$.

\item Decoding at Bob: On receiving the pair $(v(l),j)$ from
Alice and Helen, Bob declares its estimate of $X$ as the $j$th $x^{n}$-sequence belonging to the set $T^{n}_{X|v(l)}$. 
For $n$ sufficiently large, decoding at Bob will succeed with high probability.

\item Equivocation:
\begin{align}
H(X^{n}|J_x)&\geq H(X^{n}|J_x, \mathcal{E}_{H})\\
&= \sum_{j}P(J_x=j, \mathcal{E}_{H}=1)H(X^{n}|J_x=j, \mathcal{E}_{H}=1) \nonumber\\
&\quad + \sum_{j}P(J_x=j, \mathcal{E}_{H}=0)H(X^{n}|J_x=j, \mathcal{E}_{H}=0)\\
&\geq \sum_{j}P(J_x=j, \mathcal{E}_{H}=1)H(X^{n}|J_x=j, \mathcal{E}_{H}=1)\label{T2E1}.
\end{align}
Next, we note that given $J_x=j$ and $\mathcal{E}_{H}=1$, $X^{n}$ can take $2^{nR_{y}}$ values, i.e., there are a total of $2^{nR_{y}}$ $x^{n}$-sequences, each corresponding to the $j$th sequence in the (approximately) disjoint sets $T^{n}_{X|v(l)}$, for $l=1,\ldots, 2^{nR_{y}}$, and each equally likely. Therefore, we have  $P(X^{n}=x^{n}|J_x=j, \mathcal{E}_{H}=1)\approx 2^{-nR_{y}}$. Using this, we next lower bound each of the conditional entropy terms appearing in the summation of (\ref{T2E1}) as follows:
\begin{align}
&H(X^{n}|J_x=j, \mathcal{E}_{H}=1)\nonumber\\
&=\sum_{x^{n}: J_x=j, \mathcal{E}_{H}=1} P(X^{n}=x^{n}|J_x=j, \mathcal{E}_{H}=1)\log\left(\frac{1}{P(X^{n}=x^{n}|J_x=j, \mathcal{E}_{H}=1)}\right)\\
&\geq\sum_{x^{n}: J_x=j, \mathcal{E}_{H}=1} P(X^{n}=x^{n}|J_x=j, \mathcal{E}_{H}=1)\log\left(\frac{1}{2^{-n(R_{y}-\epsilon_{n})}}\right)\\
&= n(R_{y}-\epsilon_{n})\sum_{x^{n}: J_x=j, \mathcal{E}_{H}=1} P(X^{n}=x^{n}|J_x=j, \mathcal{E}_{H}=1)\\
&=n(R_{y}-\epsilon_{n})\label{T2E2}.
\end{align}
Substituting (\ref{T2E2}) in (\ref{T2E1}), we obtain
\begin{align}
H(X^{n}|J_x)&\geq \sum_{j}P(J_x=j, \mathcal{E}_{H}=1)H(X^{n}|J_x=j, \mathcal{E}_{H}=1)\\
&\geq n(R_{y}-\epsilon_{n})\sum_{j}P(J_x=j, \mathcal{E}_{H}=1)\\
&\geq n(R_{y}-\epsilon_{n})(1-\epsilon_{n})\label{T2E3}.
\end{align}
Normalizing (\ref{T2E3}) by $n$ and taking the limit $n\rightarrow \infty$, we obtain 
\begin{align}
\lim_{n\rightarrow \infty}\frac{H(X^{n}|J_x)}{n}&\geq R_{y}\\
&\geq \min(H(X), R_{y}). 
\end{align}
\end{enumerate}

\subsubsection{Converse}
The only difference in the converse part for the case of two-sided helper
is for the equivocation at the eavesdropper:
\begin{align}
H(X^{n}|J_{x})&=
H(X^{n},J_{y}|J_{x})-H(J_{y}|X^{n},J_{x})\\
&=
H(J_{y}|J_{x})-H(J_{y}|X^{n},J_{x})+H(X^{n}|J_{x},J_{y})\\
&\leq
H(J_{y}|J_{x})+n\epsilon_{n}\label{TQA}\\
&\leq
H(J_{y})+n\epsilon_{n}\\
&\leq n R_{y} + n \epsilon_{n}
\end{align}
where (\ref{TQA}) follows from Fano's inequality. Furthermore, we have the trivial upper bound $H(X^{n}|J_{x})\leq H(X^{n})=nH(X)$.
This implies the desired bound for equivocation:
\begin{align}
\Delta&\leq \min(H(X),R_{y}).
\end{align}

\subsection{Proofs of Theorems  \ref{Theorem1B} and \ref{Theorem3}}
\subsubsection{Converse Proofs}
The proofs for lower bounds on $R_{x}$ and $R_{y}$ for both Theorems \ref{Theorem1B} and \ref{Theorem3} are the same and we present these jointly. Later in this section, we present separate proofs for equivocation for each of the theorems. 

Let the coded output of the helper be denoted as $J_{y}$, and the
output of Alice be denoted as $J_{x}$, i.e.,
\begin{align}
J_{y}=f_{y}(Y^{n}),\qquad \mbox{and} \qquad J_{x}=f_{x}(X^{n},J_{y}).
\end{align}
First note that, for noiseless reconstruction of the sequence $X^{n}$ at Bob, we have by Fano's inequality
\begin{align}
H(X^{n}|J_{x},J_{y},W^{n})&\leq n\epsilon_{n}\label{fano2L}
\end{align}
We start by obtaining a lower bound on $R_{x}$, the rate of
Alice, as follows,
\begin{align}
nR_{x}&\geq H(J_{x})\\
&\geq H(J_{x}|J_{y},W^{n})\\
&= H(X^{n},J_{x}|J_{y},W^{n})-H(X^{n}|J_{x},J_{y},W^{n})\\
&\geq H(X^{n},J_{x}|J_{y},W^{n})-n\epsilon_{n}\label{eX1}\\
&\geq H(X^{n}|J_{y},W^{n})-n\epsilon_{n}\\
&=\sum_{i=1}^{n}H(X_{i}|X^{i-1},J_{y},W^{n})-n\epsilon_{n}\\
&=\sum_{i=1}^{n}H(X_{i}|J_{y},X^{i-1},W_{i+1}^{n},W_{i})-n\epsilon_{n}\label{eXX1}\\
&\geq \sum_{i=1}^{n}H(X_{i}|J_{y},Y^{i-1},X^{i-1},W_{i+1}^{n},W_{i})-n\epsilon_{n}\\
&=\sum_{i=1}^{n}H(X_{i}|V_{i},W_{i})-n\epsilon_{n}\label{eX2}\\
&=nH(X|V,W)-n\epsilon_{n}
\end{align}
where (\ref{eX1}) follows by (\ref{fano2L}) and (\ref{eXX1}) follows
from the following Markov chain,
\begin{align}
W^{i-1}\rightarrow (J_{y},X^{i-1},W_{i+1}^{n},W_{i})\rightarrow X_{i},
\end{align}
and in (\ref{eX2}), we have defined
\begin{align}
V_{i}&\triangleq(J_{y},Y^{i-1},X^{i-1},W_{i+1}^{n}).
\end{align}
We next obtain a lower bound on $R_{y}$:
\begin{align}
nR_{y}&\geq H(J_{y})\\
&\geq H(J_{y}|W^{n})\\
&\geq I(Y^{n};J_{y}|W^{n})\\
&= \sum_{i=1}^{n}H(Y_{i}|W_{i}) - H(Y^{n}|J_{y},W^{n})\\
&=\sum_{i=1}^{n}H(Y_{i}|W_{i})-\sum_{i=1}^{n}H(Y_{i}|W_{i},J_{y}, Y^{i-1},W_{i+1}^{n},W^{i-1})\\
&=\sum_{i=1}^{n}H(Y_{i}|W_{i})-\sum_{i=1}^{n}H(Y_{i}|W_{i},J_{y}, Y^{i-1},W_{i+1}^{n})\label{tempc}\\
&=\sum_{i=1}^{n}H(Y_{i}|W_{i})-\sum_{i=1}^{n}H(Y_{i}|W_{i},J_{y}, Y^{i-1},X^{i-1},W_{i+1}^{n})\label{tempd}\\
&= \sum_{i=1}^{n}H(Y_{i}|W_{i})-\sum_{i=1}^{n}H(Y_{i}|W_{i},V_{i})\\
&=nI(Y;V|W)
\end{align}
where in (\ref{tempc}) and (\ref{tempd}), we have used the Markov chain
\begin{align}
(Y_{i},J_{y},W_{i})\rightarrow Y^{i-1}\rightarrow (X^{i-1},W^{i-1}),
\end{align}
which follows from the fact that the sources $\{X_{i},Y_{i},W_{i}\}_{i=1}^{n}$ are generated i.i.d., and $J_{y}$ is a function of $Y^{n}$. 
\begin{itemize}
\item Equivocation: one-sided helper

We have the following sequence of upper bounds on the equivocation rate of the eavesdropper:
\begin{align}
H(X^{n}|J_{x})&=
H(X^{n},J_{y},W^{n}|J_{x})-H(J_{y},W^{n}|X^{n},J_{x})\\
&=H(J_{y},W^{n}|J_{x})-H(J_{y},W^{n}|X^{n},J_{x})+H(X^{n}|J_{x},J_{y},W^{n})\\
&\leq H(J_{y},W^{n}|J_{x})-H(J_{y},W^{n}|X^{n},J_{x})+n\epsilon_{n}\label{ttempa}\\
&= H(J_{y},W^{n}|J_{x})-H(J_{y},W^{n}|X^{n})+n\epsilon_{n}\label{ttempb}\\
&\leq H(J_{y},W^{n})-H(J_{y},W^{n}|X^{n})+n\epsilon_{n}\\
&= I(X^{n};J_{y},W^{n})+n\epsilon_{n}\\
&= \sum_{i=1}^{n}I(X_{i};J_{y},W^{n}|X^{i-1})+n\epsilon_{n}\\
&= \sum_{i=1}^{n}I(X_{i};J_{y},W^{n},X^{i-1})+n\epsilon_{n}\\
&\leq \sum_{i=1}^{n}I(X_{i};J_{y},W^{n},X^{i-1},Y^{i-1})+n\epsilon_{n}\\
&= \sum_{i=1}^{n}I(X_{i};W_{i},J_{y},W_{i+1}^{n},X^{i-1},Y^{i-1})+n\epsilon_{n}\\
&= \sum_{i=1}^{n}I(X_{i};W_{i},V_{i})+n\epsilon_{n}\\
&= nI(X;W,V)+n\epsilon_{n}.
\end{align}

\item Equivocation: two-sided helper

We have the following sequence of upper bounds on the equivocation rate of the eavesdropper:
\begin{align}
H(X^{n}|J_{x})&=
H(X^{n},J_{y},W^{n}|J_{x})-H(J_{y},W^{n}|X^{n},J_{x})\\
&=H(J_{y},W^{n}|J_{x})-H(J_{y},W^{n}|X^{n},J_{x})+H(X^{n}|J_{x},J_{y},W^{n})\\
&\leq
H(J_{y})+H(W^{n})- H(W^{n}|X^{n},J_{x})+n\epsilon_{n}\label{tempa}\\
&=
H(J_{y})+H(W^{n})- H(W^{n}|X^{n})+n\epsilon_{n}\label{tempb}\\
&=
H(J_{y})+nI(X;W)+ n\epsilon_{n}\\
&\leq n (R_{y}+I(X;W)) + n \epsilon_{n}
\end{align}
where (\ref{tempa}) follows from (\ref{fano2L}), and (\ref{tempb}) follows from the fact that $Y^{n}\rightarrow X^{n}\rightarrow W^{n}$,
and hence $J_{x}\rightarrow X^{n}\rightarrow W^{n}$, since $J_{x}$ is a function of $(X^{n},J_{y})$.

Furthermore, we have the trivial upper bound $H(X^{n}|J_{x})\leq H(X^{n})=nH(X)$.
This implies the desired bound for equivocation:
\begin{align}
\Delta&\leq \min(H(X),R_{y}+I(X;W)).
\end{align}
\end{itemize}

\subsubsection{Achievability}
\begin{itemize}
\item Achievability for two-sided Helper

The achievability proof for Theorem \ref{Theorem3} closely follows that of Theorem \ref{Theorem2}.
\begin{enumerate}
\item Encoding at Helen: As in the proof for Theorem \ref{Theorem2}, Helen generates $2^{nI(V;Y)}$ i.i.d. sequences, $v(l)$ from the distribution $p(v)$. Next, she independently bins these sequences in $2^{nI(Y;V|W)}$ bins; and enumerates these bin indices as $b_{v}=1,2,\ldots, 2^{nI(Y;V|W)}$. Upon observing $y^{n}$, she searches for a $v(l)$ such that $(v(l),y^{n})$ are joint typical. If successful, it transmits the bin-index of the chosen $v$-sequence. The number of sequences in each bin is approximately $2^{nI(V;W)}$ and thus upon receiving the bin-index $B(V)$ from Helen, Bob can correctly decode the $v$-sequence (using joint typical decoding). Also, since $Y\rightarrow X\rightarrow W$, we have $I(V;W)\leq I(V;X)$, and hence Alice can also correctly decode the $v$-sequence. As in the previous section, we denote $\mathcal{E}_{H}=1$ as the event that Helen's encoding is successful, the probability of which can be made arbitrarily close to $1$ by making $n$ sufficiently large and by choosing $R_{y}\geq I(Y;V)-I(Y;W)= I(Y;V|W)$. 

\item Encoding at Alice: Given that $\mathcal{E}_{H}=1$, a random $X^{n}$ sequence will belong to the conditional typical set $T^{n}_{X|\hat{v}(l)}$, where $\hat{v}(l)$ is the $v$-sequence that Alice decodes upon receiving the bin-index $B(V)$. Alice further bins the set of $x$-sequences belonging to $T^{n}_{X|\hat{v}(l)}$ into $2^{nH(X|W,V)}$ bins and denotes these as $b_{x}=1,\ldots, 2^{nH(X|W,V)}$; so that the number of $x$-sequences in each bin is approximately $2^{nI(X;W|V)}$. Alice sends the bin-index $B(X)$ in which the observed $x^{n}$-sequence falls corresponding to the conditionally typical set $T^{n}_{X|\hat{v}(l)}$. The total rate required by Alice is therefore  $H(X|W,V)$.

\item Decoding at Bob: Upon receiving $B(V)$ from Helen and $B(X)$ from Alice, Bob first decodes $v$ by searching for a unique $\hat{v}\in B(V)$ such that $(\hat{v},w^{n})$ are joint typical. The probability of decoding error in estimating $v$ at Bob goes to $0$ as $n\rightarrow \infty$ since the number of $v$ sequences in each bin is approximately $2^{nI(V;W)}$. Bob then looks in the $B(X)$th bin in the set $T^{n}_{X|\hat{v}}$; and searches for a unique $\hat{x}^{n}$ in this set such that 
$(\hat{x}^{n}, \hat{v}, w^{n})$ are joint typical. This step will lead to a successful decoding at Bob since the number of $x$-sequences in each such bin is approximately $2^{nI(X;W|V)}$.  

\item Equivocation: As in the proof for Theorem \ref{Theorem2}, we follow the same sequence of lower bounds to arrive at:
\begin{align}
H(X^{n}|B(X))&\geq \sum_{j}P(B(X)=j, \mathcal{E}_{H}=1)H(X^{n}|B(X)=j, \mathcal{E}_{H}=1)\label{TT1}
\end{align}
We next note that conditioned on the event $\mathcal{E}_{H}=1$, and given $B(X)=j$, 
there are a total of $2^{nI(X;W|V)}$ sequences in each of the bins; and each bin could have resulted from any of the $2^{n(R_{y}+I(W;V))}$ $v$-sequences. Thus, there are a total of $2^{n(R_{y}+I(W;V)+I(X;W|V))}=2^{n(R_{y}+I(X;W))}$ equally likely $x^{n}$-sequences conditioned on $B(X)=j$ and $\mathcal{E}_{H}=1$. We therefore have $P(X^{n}=x^{n}|B(X)=j, \mathcal{E}_{H}=1)\approx 2^{-n(R_{y}+I(X;W))}$. 
Using this, we can bound
\begin{align}
H(X^{n}|B(X)=j, \mathcal{E}_{H}=1)&\geq n(R_{y}+I(X;W)-\epsilon_{n})\label{TT2}
\end{align}
Upon substituting (\ref{TT2}) in (\ref{TT1}), and letting $n\rightarrow \infty$, we obtain at the resulting equivocation of this scheme as:
\begin{align}
\lim_{n\rightarrow \infty}\frac{H(X^{n}|B(X))}{n}&\geq R_{y}+I(X;W)\\
&\geq \min(H(X), R_{y}+I(X;W)). 
\end{align}

\end{enumerate}

\item Achievability for one-sided Helper. Encoding at Helen remains the same as the two-sided helper case, i.e., Helen quantizes $Y^{n}$ to $V^{n}$ and performs binning with respect to $W^{n}$. The encoding at Alice is to independently and uniformly bin the set of $X$-sequences in $2^{nH(X|W,V)}$ bins
and it sends the bin index $B(X^{n})$. The only difference is in the equivocation proof:
\begin{align}
H(X^{n}|B(X^{n}))&=  H(X^{n}) - I(X^{n};B(X^{n})) \\
&= nH(X) - H(B(X^{n})) + H(B(X^{n})|X^{n})\\
&=nH(X) - H(B(X^{n})) \label{Tyu}\\
&\geq nH(X) - \log(|B(X^{n})|) \\
&\geq nH(X) - \log(2^{nH(X|W,V)}) \\
&= nI(X;W,V).
\end{align}
where in (\ref{Tyu}), we used the fact that $B(X^{n})$ is a deterministic function of $X^{n}$. We therefore have
\begin{align}
\lim_{n\rightarrow \infty}\frac{H(X^{n}|B(X))}{n}&\geq I(X;W,V).
\end{align}

\end{itemize}

\bibliographystyle{unsrt}
\bibliography{refravi}
\end{document}